\documentclass[smallextended]{svjour3}   
\usepackage{graphicx}                                   
\usepackage{mathptmx}                                   
\usepackage{amsmath}                                            
\usepackage{amsfonts}                                   
\usepackage{amssymb}                                            
\usepackage[caption=false]{subfig}                                              
\usepackage{marvosym}
\usepackage{ifpdf}                                              

\ifpdf
\usepackage[unicode,colorlinks,linkcolor=blue,urlcolor=blue,citecolor=blue,
plainpages=false,pdfpagelabels,breaklinks]{hyperref}
\else
\usepackage[unicode,colorlinks,linkcolor=blue,urlcolor=blue,citecolor=blue,
plainpages=false,pdfpagelabels,linktocpage]{hyperref}
\fi

\usepackage[numbers,square,sort&compress]{natbib}

\newcommand{\be}{\begin{equation}}
\newcommand{\ee}{\end{equation}}
\newcommand{\bea}{\begin{eqnarray}}
\newcommand{\eea}{\end{eqnarray}}

\def\gsim{\mathrel{\raise.5ex\hbox{$>$}\mkern-14mu
             \lower0.6ex\hbox{$\sim$}}}
\def\lsim{\mathrel{\raise.3ex\hbox{$<$}\mkern-14mu
             \lower0.6ex\hbox{$\sim$}}}

\def\beq{\begin{equation}}
\def\eeq{\end{equation}}

\begin{document}

\title{Radiation from charged particles on eccentric orbits in a dipolar magnetic field around a Schwarzschild black
hole}

\titlerunning{Radiation from charged particles on eccentric orbits} 

\author{D. B. Papadopoulos \and I. Contopoulos  \and K. D. Kokkotas \and N. Stergioulas}

\authorrunning{D. B. Papadopoulos et al.} 

\institute{D. B. Papadopoulos$^1$ \and I. Contopoulos$^2$ \and K.
D. Kokkotas$^3$ \and N. Stergioulas$^1$ \at
              $^1$Department of Physics, Aristotle University of Thessaloniki, Thessaloniki 54124, Greece \\
              $^2$Research Center for Astronomy and Applied Mathematics, Academy of Athens, Athens 11527, Greece\\
              $^3$Theoretical Astrophysics, Eberhard Karls University of T\"{u}bingen, T\"{u}bingen 72076, Germany}

\date{Received: date / Accepted: date}
\maketitle

\begin{abstract}
We obtain an approximate solution for the motion of a charged
particle around a Schwarzschild black hole immersed in a weak
dipolar magnetic field. We focus on eccentric bound orbits in the
equatorial plane of the Schwarzschild black hole and derive an analytic expression for the spectral distribution of the electromagnetic emission from a charged particle on such an orbit.
Two sets of harmonic contributions appear, with specific frequency spacing. The expression can be written in compact form, if it is truncated up to the lowest order harmonic contributions.\end{abstract}



\section{Introduction}

 The origin of astrophysical magnetic fields on galactic and
extragalactic scales remains one of the open questions of modern
astrophysics. Microgauss magnetic fields have been observed in
elliptical and spiral galaxies, in the intracluster medium, and in
damped Lyman alpha systems over cosmological distances (see
e.g.~\cite{w2} for a review). In the past, several investigations
of bremsstrahlung radiation were carried out in the case of a
Schwarzschild black hole embedded in a large-scale homogeneous
asymptotically uniform external magnetic field
(\cite{G1,G2,G3}).The total radiated power was obtained through a
generalized Larmor formula based on arguments about covariance
under Lorentz transformations (\cite{a3a,a3}). In a broad sense,
bremsstrahlung~\cite{w2a} is the radiation emitted by an
accelerated charged particle. The angular distribution of the
radiation produced from an orbiting charged particle in the
presence of a uniform magnetic field has been discussed
extensively with the aid of the Lienard-Wiechert potentials at
large distances in~\cite{a3,a4,a4a}. Along the way, several
interesting features of the dynamics of charged particles in the
presence of both the black hole gravitational field and the
uniform magnetic field were investigated. In~\cite{p0} and
references therein, the authors explored two different types of
orbits, with and without curls. This analysis was generalized
by~\cite{p1} in the case of a Schwarzschild black hole embedded in
a dipole magnetic field. An interesting result obtained
in~\cite{p2} was the possibility of trapping relativistic charged
particles in a potential well inside the position of the innermost
stable circular orbit. In the present work, we investigate the
eccentric orbits of charged particles around a Schwarzschild black
hole embedded in a dipole field and obtain their electromagnetic
radiation spectrum, assuming slow motions and weak fields. Our
paper is organized as follows: In Section 2 we present the
mathematical formalism. In Section 3, we investigate eccentric
bound orbits in the equatorial plane and in Section 4, we discuss
the angular distribution of the electromagnetic radiation from
those particles. We close with a discussion in Section 5.

\section{Equations of motion in Schwarzschild space-time}

We follow the $3+1$~formalism of ~\cite{a3a} and ~\cite{p2a} for a dipole magnetic field in
a Schwarzschild background and work in geometric units with $c=G=1$ (in some equations the speed of light $c$ is kept on purpose).
Greek spacetime indices take values from 0 to 3 and Latin indices from 1 to 3.  In a
$(t,r,\theta,\phi$) coordinate system the Schwarzschild metric
is
\begin{equation}\label{s1}
ds^2=-\left(1-\frac{r_S}{r}\right)dt^2+\frac{dr^2}{1-\frac{r_S}{r}}+r^2
\left(d\theta^2+\sin^2{\theta}d\phi^2\right),
\end{equation}
where $r_S=2M$ is the Schwarzschild radius of the black hole. The
components of the covariant four-velocity are $u^{\mu}=
(u^0,u^{i})$, with $u_\mu u^\mu=-1$.
%
%
The equations of motion of a charged test particle in the presence
of a dipole magnetic field can be derived from the Lagrangian
\begin{equation}\label{e1}
2{\cal
L}=g_{\mu\nu}\frac{dx^\mu}{d\tau}\frac{dx^\nu}{d\tau}+\frac{q}{m}\frac{dx^\mu}{d\tau}A_{\mu},
\end{equation}
where $\tau$ is the
proper time and $m$ and $q$ are the particle's mass and electric
charge respectively. The Lagrangian now reads
\begin{eqnarray}\label{e2}
2{\cal
L}&=&-\left(1-\frac{r_S}{r}\right)\dot{t}^2+\frac{\dot{r}^2}{1-\frac{r_S}{r}}+r^2
\left(\dot{\theta}^2+\sin^2{\theta}\dot{\phi}^2\right)
+\frac{q}{m}\frac{dx^\mu}{d\tau}A_{\mu}.
\end{eqnarray}
where a dot denotes differentiation with respect to $\tau$.

The motion of a charged particle around a Schwarzschild black hole
immersed in a dipole magnetic field~\cite{p1},~\cite{p2} is
characterized by two conserved quantities, the energy $E>0$ and the
generalized angular momentum $L$, given by the expressions
\begin{eqnarray}\label{f1a}
&&\dot{t}=E\left(1-\frac{r_S}{r}\right)^{-1},\\
&&\dot{\phi}=\frac{L}{r^2\sin^2{\theta}}-\frac{q}{m}\frac{A_\phi}{r^2\sin^2{\theta}},\\
&&\dot{\theta}=\frac{p_\theta}{r^2} \, .
\end{eqnarray}
We assume a dipole electromagnetic four-potential of the form
$A_{\mu}=(0,0,0,A_{\phi})$, where
\begin{equation}\label{s5a}
A_{\phi}=\frac{3\mu_b}{r_S^3}r^2\sin^2{\theta}W(r),
\end{equation}
and
\begin{equation}\label{s5b}
W(r)\equiv \ln{\left(1-\frac{r_S}{r}\right)}+\frac{r_S}{r}  +\frac{r_S^2}{2
r^2} \, .
\end{equation}
Above, $\mu_b$ is the magnetic dipole moment for an
observer at infinity. Thus, for the metric (\ref{s1}) the normalization of the 4-velocity yields
\begin{eqnarray}\label{s13}
\dot{r}^2 & = & E^2-\left(1-\frac{r_S}{r}\right)\left[
1+r^2\dot{\theta}^2
+r^2\sin^2{\theta}\left(\frac{L}{r^2\sin^2{\theta}}
-\frac{3q\mu_b}{mr_S^3}W(r)\right)^2\right] \, .
\end{eqnarray}
We make the following changes in the above equation. First, we
change the $\tau$-derivatives to $\phi$-derivatives, through the
relation $\dot{r}=({dr}/{d\phi})\dot{\phi}$; second,  we assume
that the particle motion is confined in the equatorial plane
($\theta=\pi/2$); finally, we assume that $r\gg r_S$, thus
$W(r)\approx -{r_S^3}/{3r^3}$, and we arrive at the following
equation of motion, which describes the geometry of charged
particle orbits in the invariant equatorial plane


\begin{eqnarray}\label{s14e}
\left(\frac{du}{d\phi}\right)^2&=&\frac{1}{(1+\Omega u)^2}
\left[\frac{E^2-1}{L^2}+\frac{r_S}{L^2}u-u^2 +(r_S-2\Omega)u^3 \right . \nonumber \\
 && \ \ \ \ \ \ \ \ \ \ \ \ \ \ \ \ \ \ \ \ + \left . (2\Omega r_S-\Omega^2)u^4+\Omega^2 r_S u^5\right],
 \label{11}
\end{eqnarray}
where $u=1/r$ and
\begin{equation}
\Omega\equiv\frac{q\mu_b}{m L}\ .
\end{equation}
The parameter $\Omega$
can be negative or positive, depending on the sign of the
charge $q$ and on the azimuthal generalized angular momentum $L$.
In what follows, we will restrict ourselves to positive values of
$\Omega$.

\section{Bound orbits}

The motion of charged particles in a dipolar magnetic field around
a static black hole was studied in~\cite{p1,p2} and \cite{p2p}
(see also the review in~\cite{G1}). A synchrotron mechanism was
considered for circular orbits of charged particles in the
presence of a uniform magnetic field. Here, we focus only on
eccentric bound orbits of charged particles of the form
$r=l/(1+e\cos{\xi})$, where $e$ is the eccentricity of the orbit
with $0<e<1$, $l$ is its latus rectum and $\xi\in[0,2\pi]$ in the
equatorial plane of a Schwarzschild black hole, immersed in a
dipole magnetic field, when $r/r_S\gg 1$. We seek approximate,
analytic solutions, by keeping  only orders up to $O(u^3)$ in the
series appearing in the nominator of (\ref{11}). With theses
assumptions, we find values of $E,\ L$ and $\Omega$ such that
Eq.~(\ref{s14e}) admits bound orbits. More details on this method
can be found in  \cite{a5y,a5x, a6, a6a}. The geometry of the
orbits is determined by the real roots of the equation $f_1(u)=0$
where
\begin{equation}\label{s14d}
f_1(u)=-\frac{1-E^2}{L^2}+\frac{r_S}{L^2}u-u^2+ \left(r_S-2\Omega\right)u^3.
\end{equation}
(when  $1+\Omega u\neq 0)$. The orbital
eccentricity $e$ and latus rectum $l=a(1-e^2)$ (with $a$ being
the major axis of the ellipse) satisfy
\begin{eqnarray}\label{e6}
\frac{r_S}{L^2}&=&\frac{1}{l^2}\left[2l-(3+e^2)(r_S-2\Omega)\right],\\
\frac{(1-E^2)}{L^2}&=&\frac{1}{l^3}\left[l-2(r_S-2\Omega)\right](1-e^2)
\label{latusrectum} \, .
\end{eqnarray}
In this case, the equation $f_1(u)=0$ admits three real roots
which are
\begin{equation}\label{e7}
u_1=\frac{1-e}{l}\, , \quad u_2=\frac{1+e}{l}\, , \quad
u_3=\frac{1}{r_S-2\Omega}-\frac{2}{l}\ .
\end{equation}
From ~(\ref{e6}) we deduce that
\begin{equation}\label{mu1}
\mu\equiv\frac{M}{l}\leq\frac{1}{3+e^2}+\frac{\Omega}{l},
\end{equation}
and assuming that $u_3\geq u_2$ we arrive at another interesting inequality
\begin{equation}\label{mul1}
{r_S}-{2\Omega}\leq\frac{l}{3+e} \, .
\end{equation}
Substituting ({\ref{e6}), (\ref{latusrectum}) and
\begin{equation}\label{u1}
u=\frac{1}{l}(1+e\cos{\xi}),
\end{equation}
into (\ref{s14e}), we obtain
\begin{eqnarray}\label{e10}
\left(\frac{d\xi}{d\phi}\right)^2&=&\frac{1}{[1+(\Omega/l)(1+e\cos{\xi})]^2}
 \left[1- \left(3+e\cos{\xi}\right) \frac{r_S-2\Omega}{l}\right],
\end{eqnarray}
which reduces to
\begin{eqnarray}\label{e11}
\left[1+\frac{\Omega}{l}(1-e)\right]\frac{d\xi}{\sqrt{1-k^2\cos^2{(\xi/2)}}}
-\left(\frac{2\Omega e}{l}\right)\frac{\cos^2{(\xi/2)}d\xi}{\sqrt{1-k^2\cos^2{(\xi/2)}}}
\nonumber\\
=\pm\sqrt{\lambda}d\phi,
\end{eqnarray}
with $\lambda\equiv 1-(3-e)\frac{r_S-2\Omega}{l}> 0$ and
\begin{equation}\label{e12}
k^2=\frac{2e(r_S-2\Omega)}{l-(3-e)(r_S-2\Omega)}<1 \, .
\end{equation}
Integrating
(\ref{e11}) we find
\begin{equation}
2\left[l+{\Omega}(1-e)\right]F[\cos{(\xi/2)},k]
-{4\Omega e}\left\{F[\cos{(\xi/2)},k]-E[\cos{(\xi/2)},k]\right\}=\pm\sqrt{\lambda}l \phi,
\label{e13}
\end{equation}
where $F[\cos{(\xi/2)},k]$ and $E[\cos{(\xi/2)},k]$ are the
elliptic integrals of first and second kind, respectively, with
modulus $k<1$.

Expanding the elliptic integrals in terms of the small modulus $k$ and keeping only zero order terms
in $k<1$ (see ~\cite{a5a, a5b, a5c})
, we obtain
\begin{equation}\label{e16}
\xi-\frac{\Omega e}{l+\Omega(1-2e)}\sin{\xi}=
\pi\pm\chi \phi \, .
\end{equation}
where $\frac{\Omega e}{l+\Omega(1-2e)}<1$ and $\chi= \frac{l\sqrt{\lambda}}{l+\Omega (1-2e)}$.

From (\ref{e16}) we compute $\cos{\xi}$, which appears in
(\ref{u1}) and obtain
\begin{equation}\label{e20}
\cos{\xi}=-\cos{(\chi\phi)}
\mp\frac{\Omega e}{l+\Omega(1-2e)}
\sin^2{\left(\chi\phi\right)}.
\end{equation}
Substitution of (\ref{e20}) into (\ref{u1}) yields
\begin{equation}\label{e21}
u=\frac{1}{l}\left[ 1-e\cos\left(\chi\phi \right) \mp\frac{\Omega e^2}{l+\Omega(1-2e)} \, \sin^2\left(\chi\phi\right)\right].
\end{equation}
To linear order in $e$, the last equation becomes
\begin{equation}\label{e22}
u=\frac{1}{l}[1-e\cos{(\chi\phi)}],
\end{equation}
where the origin of $\phi$ may be determined by Eq.(\ref{e16}) setting $\xi=0$.

The approximate solution may now be completed by direct integration of the
equations
\begin{eqnarray}\label{a23}
\frac{d\phi}{d\tau} & = & L u^2(1+\Omega u)\ ,\nonumber\\
\frac{d t}{d\tau} & = & E(1-r_su)^{-1}\ .
\end{eqnarray}
From  (\ref{e22}) and (\ref{a23}) we obtain
\begin{equation}\label{a26}
\frac{d(\chi\phi)}{[1-e\cos{(\chi\phi)}]^2[\rho_1+\rho_2\cos{(\chi\phi)}]}=\frac{\chi L}{El^2} dt,
\end{equation}
where we have neglected the $O(e^2)$ term $2e^2(\frac{\mu\Omega}{l})$ and set  $\rho_1=(1-2\mu)(1+\frac{\Omega}{l})$ and
$\rho_2=2\mu-\frac{\Omega}{l}(1-4\mu)$.

Integrating
(\ref{a26}), we find
\begin{eqnarray}\label{a27}
N t &\equiv& B_1\arctan{\left(\sqrt{\frac{1+e}{1-e}}\tan{\frac{(\chi\phi)}{2}}\right)}
+B_2\arctan{\left(\sqrt{\frac{\rho_1+e\rho_2}{\rho_1-e\rho_2}}\tan{\frac{(\chi\phi)}{2}}\right)}
\nonumber\\
&+&
B_3\frac{\sin{(\chi\phi)}}{2(1-e\cos{(\chi\phi))}},
\end{eqnarray}
where
\begin{eqnarray}\label{a28}
&&B_1=\frac{2[(\rho_1+\rho_2)+\rho_2(1-e^2)]}{(1-e^2)^{3/2}(\rho_1+\rho_2)^2},\nonumber\\
&&B_2=\frac{2\rho_2^2}{(\rho_1+\rho_2)^2\sqrt{\rho_1^2-e^2\rho_2^2}},\nonumber\\
&&B_3=\frac{2e}{(1-e^2)(\rho_1+\rho_2)},
\end{eqnarray}
and
\begin{eqnarray}\label{f1}
N\equiv\frac{\chi L}{2E l^2}
=\frac{L}{2El^2}\frac{\sqrt{1-(3-e)\frac{r_S-2\Omega}{l}}}
{1+\frac{\Omega}{l}(1-2e)},
\end{eqnarray}
has dimensions of frequency. In Appendix B, we compute the
right-hand of (\ref{f1}) in terms of $e$, $\mu$ and $\Omega\neq
0$ with the aid of (\ref{e6}), (\ref{latusrectum}) and find a rather lengthy
expression. However, for $\Omega=0$, we compute the ratio
${L}/{E}$ using (\ref{e6}) and recover the Newtonian
frequency of a Keplerian orbit with the same $e$ and $l$.

Note that, when both the eccentricity $e\in(0,1)$ and
$\mu=M/l=r_S/(2l)$  are small, the coefficients $B_1$,
$B_2$ and $B_3$ can be approximated as
\begin{equation}\label{a30}
B_1=2+O(e)\sim 2 \, , \quad B_2= O(\mu^2)\, , \quad B_3= O(e) \, .
\end{equation}
Thus, (\ref{a27}) takes the approximate form
\begin{equation}\label{a31}
\tan\left(\frac{\chi\phi}{2}\right) \simeq\sqrt{\frac{1-e}{1+e}}\tan{(N
t)},
\end{equation}
since the second and the third term are very small in comparison to the first one multiplied by $B_1$. Furthermore, using (\ref{a31}) we find
\begin{eqnarray}\label{a32}
&&\cos{(\chi \phi)}=\frac{\cos{(2Nt)}-e}{1-e\cos{(2Nt)}},\nonumber\\
&&\sin{(\chi \phi)}=\sqrt{1-e^2}\left(\frac{\sin{(2Nt)}}{1-e\cos{(2Nt)}}\right).
\end{eqnarray}
Finally, because of our previous approximations in
(\ref{a30}), Eq.~(\ref{e22}) yields
\begin{equation}\label{a33}
r=\frac{l[1-e\cos{(2 N t)}]}{(1+e^2)-2e\cos{(2Nt)}} \, .
\end{equation}

\section{Spectral distribution of radiation}

We are next interested in examining the electromagnetic emission
of a particle with charge $q$ that is following an eccentric bound
orbit, such as the one discussed in Section III. Our aim is to
obtain analytic expressions for the distribution of energy as a
function of  angle and radiation frequency (see e.g. \cite{a3, a4,
a4a}), valid under the assumptions made in Section III. We make
use of the well-known result \cite{a3, a4, a4a}
\begin{equation}\label{se17}
\frac{d^2 {\cal E}}{d\omega' d\Omega_s}=\frac{q^2\omega'^2}{4\pi^2
} \left| \int_{-\infty}^{\infty} {\bf n}\times ({\bf n}\times {\rm \bf
v})e^{i\omega'(t-{\bf n} \cdot {\bf r}(t))} dt\right|^2\ ,
\end{equation}
where ${\cal E}$ is the total energy radiated per unit frequency
$\omega'$ per unit solid angle $\Omega_s$ \cite{a4}  in which we
replace the internal and external vector products with
general-relativistic expressions  (derived in the
$3+1$~formalism~\cite{p2,p2a}).  For our further derivations, it
is necessary to know the spatial velocity $ {\rm \bf v}$ and
position ${\bf r}(t)$ over a small arc of the trajectory whose
tangent is pointing towards the observation point. In our case the
segment of the trajectory lies in the $xy$-plane, and the unit
vector ${ {\rm \bf n}}$ in the direction of radiation lies
(without loss of generality) in the $yz$-plane at an angle
$\lambda$ with the $z$-axis. In this case, the unit vector ${\bf
{\rm \bf n}}$ is
\begin{equation}\label{se18}
{\bf n}=\sin{\lambda} {\bf e}_y+\cos{\lambda} {\bf e}_z,
\end{equation}
where ${\bf e}_y$,   and ${\bf e}_z$ are the unit vectors along
the $y$ and $z$ axes, respectively. In \cite{a3,a4} the authors examined the
angular distribution of the energy radiated per unit frequency
$\omega'$ per unit solid angle $\Omega_s$  for $\lambda\approx
\pi/2$, since appreciable radiation intensity is
concentrated mainly in the equatorial plane of the orbit, and the duration of the pulse
is short. Here, we find an expression   for  any angle $\lambda\in(0,2\pi)$ (but which is valid only under the assumptions mentioned in Section III).

The internal and external vector products in (\ref{se17}) are derived in
spherical polar coordinates and become \footnote{The cross vector product is
$({\bf n}\times {\bf {\rm v}})_{l}=\varepsilon_{lkm} n^k {\rm v}^m
$ with
$\varepsilon_{123}=\frac{r^2\sin{\theta}}{\sqrt{(1-r_S/r)}}$,
while in the dot  product  $\gamma_{ij}$ is the spatial
part of the metric.}
\begin{eqnarray}\label{se20a}
{\bf {\rm} n}\cdot{\bf r}=\gamma_{ij} n^i r^j
=\frac{r}{\sqrt{1-r_S/r}}\left[\sin{\lambda}\sin{\theta}\sin{\phi}+\cos{\theta}\cos{\lambda}\right],
\end{eqnarray}
and
\begin{equation}\label{se19}
{\bf n}\times ({\bf n}\times {\rm {\bf v}})=X{\bf
e}_r+\frac{Y}{r}{\bf e}_{\theta}+\frac{Z}{r\sin{\theta}}{\bf
e}_{\phi},
\end{equation}
 where the explicit expressions of $X,Y$
and $Z$ are given in Appendix A.
In the equatorial plane ($\theta=\pi/2$), (\ref{se17}) then becomes:
\begin{eqnarray}\label{se21b}
\frac{d^2 {\cal E}}{d\omega' d\Omega_s}
=\frac{q^2\omega'^2}{4\pi^2 }\left|\int_{-\infty}^{\infty} \left[X_0{\bf e}_r+\frac{Y_0}{r}{\bf e}_{\theta}+\frac{Z_0}{r}{\bf e}_{\phi}\right]e^{i\omega' t}
\times\exp{\left \{ -i\omega'\frac{r\sin{\lambda}\sin{\phi}}{\sqrt{(1-r_S/r)}} \right \}} dt\right|^2,
\nonumber
&& \\
\end{eqnarray}
where
\begin{eqnarray}\label{se22}
X_0&\equiv&-\frac{1}{\sqrt{(1-r_S/r)}}\left[\cos^2{\lambda}\frac{d r}{dt}
-\sin^2{\lambda}\cos{\phi}\frac{d(r\cos{\phi})}{dt}\right],\\
Y_0&\equiv&- r\sin{\lambda}\cos{\lambda}\frac{d(r\sin{\phi})}{dt},\\
Z_0&\equiv&-r \left[\sin^2{\lambda}\sin{\phi}\frac{d(r\cos{\phi})}{dt}-\cos^2{\lambda} \left(r\frac{d\phi}{dt}\right)\right].
\end{eqnarray}
Using (\ref{sa4}), the exponential function in
(\ref{se21b}) becomes
\begin{eqnarray}\label{ap22}
e^{i\omega'(t-{\bf n} \cdot {\bf r}(t))}&=&\sum_{m=-\infty}^{\infty} J_{m}(Q_1)e^{i t(\omega'-2m N)}
+\sum_{m=-\infty}^{\infty} J_{m}(Q_2) e^{it (\omega'-4 m N) },
\end{eqnarray}
where $J_m$ are the Bessel functions of the first kind, while we set
\begin{eqnarray}\label{q1}
Q_1=\omega'\sin{\lambda}\frac{l}{\sqrt{1-2\mu}} \quad \mbox{and} \quad
Q_2=\omega' \sin{\lambda}\frac{l(2-5\mu)e}{2 (1-2\mu)^{3/2}}.
\end{eqnarray}
Eq.~(\ref{se21b}) with the help of (\ref{sa1}) and
(\ref{ap22}) now reads
\begin{equation}\label{sa5}
\frac{d^2 {\cal E}}{d\omega' d\Omega_s}=\frac{q^2\omega'^2}{4\pi^2
}\left[(W_1^m)^2+(W_2^m)^2+(W_3^m)^2\right],
\end{equation}
where
\begin{eqnarray}\label{sa6}
W_1^m\equiv\int_{-\infty}^{\infty}e^{i\omega'(t-{\bf n} \cdot {\bf r}(t))} X_0 dt &=&2i\sqrt{\frac{\pi}{2}}l N\left(\sum_{m=-\infty}^{+\infty}\delta(\omega'-4N m)W_{1b}^m \right . \nonumber\\
&& \ \ \ \ \ \ \ \ \ \ \ \ \ \ \ \ \ \ \ \ \left . -\sum_{m=-\infty}^{+\infty}\delta(\omega'-2N m)W_{1a}^m
\right),
\label{sa7} \\
W_2^m\equiv\int_{-\infty}^{\infty}e^{i\omega'(t-{\bf n} \cdot {\bf r}(t))} \frac{Y_0}{r} dt
&=&-\sqrt{\frac{\pi}{2}}l N \sin{2\lambda}\left(\sum_{-\infty}^{\infty}\delta(\omega'-2m N )W_{2a}^m \right .\nonumber \\
&& \ \ \ \ \ \ \ \ \ \ \ \ \ \ \ \ \ \ \ \ \ \ \ \ \ \ \ \ \ \left . +\sum_{-\infty}^{\infty}\delta(\omega'-4m N )W_{2b}^m\right),\\ \label{sa8}
W_3^m\equiv\int_{-\infty}^{\infty}e^{i\omega'(t-{\bf n} \cdot {\bf r}(t))} \frac{Z_0}{r} dt
&=& -2\sqrt{\frac{\pi}{2}}l N \left(\sum_{-\infty}^{\infty}\delta(\omega'-2m N )W_{3a}^m \right . \nonumber \\
&& \left . \ \ \ \ \ \ \ \ \ \ \ \ \ \ \ \ \ \ \ \ \ + \sum_{-\infty}^{\infty}\delta(\omega'-4m N )W_{3b}^m\right).
\end{eqnarray}
Furthermore, (\ref{sa5}) with the aid of (\ref{sa6}),  (\ref{sa7}) and (\ref{sa8}) yield
\begin{eqnarray}\label{sa9}
\frac{d^2 {\cal E}}{d\omega' d\Omega_s}&=& \frac{T}{4\pi^2}
q^2\omega'^2 l^2 N^2 \nonumber\\
&&\times \left \{\sum_{m=-\infty}^{+\infty}
\delta(\omega'-2N m)\left[(W_{1a}^m)^2+(W_{2a}^m)^2+(W_{3a}^m)^2\right] \right .\nonumber \\
&& \left .  \ \ \ \ +\sum_{m=-\infty}^{+\infty}\delta(\omega'-4N m)\left[(W_{1b}^m)^2+
(W_{2b}^m)^2+(W_{3b}^m)^2 \right] \right \} .\label{48}
\end{eqnarray}
We have used the approximation that $\delta^2(\omega')=
\frac{1}{2\pi/T}\delta(\omega')$ in the limit $T\rightarrow
\infty$, where $T$ is the radiation emission time during the
orbital motion of the charged particle around the black hole, while we set
\begin{eqnarray}
W_{1a}^m & = & 2\cos^2{\lambda} (1-2\mu)^{1/4} e  \left[J_{m+1}(Q_1)-J_{m-1}(Q_1)\right] \nonumber\\
&&-\sin^2{\lambda} \left \{\frac{(1-2\mu)^2}{2}
\left[J_{m+2}(Q_1)-J_{m-2}(Q_1)\right] \right . \nonumber\\
&&\ \ \ \ \ \ \ \ \ \ \ \ \ \ \ -2(1-2\mu)^2 e
\left[J_{m+1}(Q_1)-J_{m-1}(Q_1)\right]  \nonumber\\
&& \left .+\frac{(1-2\mu)(7-6\mu)e}{8}
\left[J_{m+3}(Q_1)-J_{m-3}(Q_1) +J_{m+1}(Q_1)-J_{m-1}(Q_1)\right] \right \},
\label{m1}\\
W_{1b}^m & = & \cos^2{\lambda}(1-2\mu)^{1/4} e
[J_{2m+1}(Q_2)-J_{2m-1}(Q_2)] \nonumber\\
&&- \sin^2{\lambda}\left \{\frac{(1-2\mu)^2}{2}
[J_{2m+2}(Q_2)-J_{2m-2}(Q_2)] \right .\nonumber \\
&&-2(1-2\mu)^2 e
[J_{2m+1}(Q_2)-J_{2m-1}(Q_2)] \nonumber\\
&& \left .+
\frac{(1-2\mu)(7-6\mu)e}{8}
[J_{2m+3}(Q_2)-J_{2m-3}(Q_2)
 +J_{2m+1}(Q_2)-J_{2m-1}(Q_2)] \right\}, \nonumber \\
 \\
W_{2a}^m & = & [J_{m-1}(Q_1)+J_{m+1}(Q_1)]
+2e[J_{m-2}(Q_1)+J_{m+2}(Q_1)],\\ \label{m2}
W_{2b}^m & = & [J_{2m-1}(Q_2)+J_{2m+1}(Q_2)]
+2e[J_{2m-2}(Q_2)+J_{2m+2}(Q_2)], \\
W_{3a}^m & = & \sin^2{\lambda} \left\{\frac{1}{2}[2J_m(Q_1)
+J_{m-2}(Q_1)-J_{m+2}(Q_1)] \right .\nonumber\\
&& \left .+\frac{5e}{4}[J_{m-1}(Q_1)+J_{m+1}(Q_1)
-J_{m-3}(Q_1)-J_{m+3}(Q_1)] \right \}\nonumber\\
&&+2e\cos^2{\lambda}
[J_m(Q_1)+J_{m-1}(Q_1)+J_{m+1}(Q_1)],\\ \label{m3}
W_{3b}^m & = & \sin^2{\lambda} \left \{\frac{1}{2}[2J_m(Q_2) +J_{2m-2}(Q_2)-J_{2m+2}(Q_2)] \right .\nonumber\\
&& \left . +\frac{5e}{4}[J_{2m-1}(Q_2)+J_{2m+1}(Q_2) -J_{2m-3}(Q_2)-J_{2m+3}(Q_2)] \right \}\nonumber\\
&&+ \frac{e}{2}\cos^2{\lambda}[4J_m(Q_2)+J_{2m-1}(Q_2)+J_{2m+1}(Q_2)
J_{2m+3}(Q_2)-J_{2m-3}(Q_2)].
\end{eqnarray}
Recall, that the argument $Q_2$ depends on the geometry of the orbit, since it is proportional to the eccentricity.
The analytic expression (\ref{48}) is our final result.
\section{Discussion}

We obtained an analytic expression for the spectral distribution
of the electromagnetic emission from charged particles in an
eccentric, bound equatorial orbit around a Schwarzschild black
hole, embedded in a dipolar magnetic field. Our final result
(\ref{sa9}) includes an infinite number of harmonics. We note that
in the second sum, the argument $Q_2$ in the Bessel functions
depends on the eccentricity of the charged particle, but in the
first sum, the argument $Q_1$  does not. It is evident from the
analytic expression~(\ref{sa9}) that the spectrum of the emitted
radiation is composed of spectral lines appearing at frequencies
intervals $\Delta\omega_1=2N$ and $\Delta\omega_2=4 N$, where $N$
is given by (\ref{f1}).

Our main result (\ref{sa9}) can be further integrated over all
frequencies $\omega'$ and all angles, yielding a distribution in
terms of the harmonics $m$. The relative contribution of
individual harmonics will depend on the precise orbit. For small
eccentricities, such as those considered here, we expect only the
first few harmonics to be important. Under this assumption, one\
could thus truncate the infinite series in (\ref{sa9}), in order
to arrive at a more compact analytic expression. We can obtain a
rough estimate of the total radiated power for an electron in
orbit around a ten solar mass Schwarzschild black hole with
$M=2\times 10^{34}$~g, $r_S= 3\times 10^6$~cm, immersed in a
dipole magnetic field typical of intergalactic space of order
$B\sim 1 \mu{\rm G}$. We further restrict our analysis to
equatorial orbits with eccentricity $e=0.5$ and latus rectum on
the order of $l\sim r_{\rm ISCO}$. In that case, $L=0.53\times
10^7\mbox{cm}^2/\mbox{sec}$, $\Omega\equiv \frac{q \mu_b}{m
L}\simeq 0.81\times 10^5\ {\rm cm}$, $\mu=r_S/2l=1/6$, and
$\frac{\Omega}{cl}\simeq 1.41\times 10^{-3}$ is indeed less than
1, so our approximations are valid. We obtain an estimate of the
ratio $L/El^2\simeq \pm 0.79 \times 10^7 {\rm sec/cm^2}$, and $N
l/c\simeq 0.141$. The total radiated power ${\cal P}$ is then
obtained by integrating Eq.~(\ref{sa9}) over all frequencies and
solid angles, and by dividing with $T$, namely ${\cal P}\simeq
1.4\times 10^{-23}~\mbox{erg/sec}$. This yields a timescale for
orbit evolution on the order of $\tau\sim m_e v^2/{\cal P}\sim
10^{16}$~sec, which is dynamically unimportant.
In a future publication, we will relax our current working
assumption that $\frac{\Omega}{cl}\ll 1$, and will also consider
higher values of the magnetic field for which radiation effects
will become astrophysically significant.

\begin{acknowledgements}
We are grateful to Toni Font for useful comments. One of us
(D.B.P) is grateful for the hospitality of the Theoretical
Astrophysics group at the University of Tuebingen, where part of
the work was done. This work was supported, in part, by the
General Secretariat for Research and Technology of Greece and by
the European Social Fund in the framework of Action `Excellence'.
\end{acknowledgements}
\section*{Appendix A}

The explicit expressions for $X,Y,Z$ are
\begin{eqnarray}\label{se20}
&&X\equiv(\varepsilon_{123})^2[g^{33}n^2(n^1 {\rm v}^2-n^2 {\rm v}^1)-g^{22}n^3(n^3 {\rm v}^1-n^1 {\rm v}^3)],\\
&&Y\equiv(\varepsilon_{123})^2 [-g^{33}n^1(n^1 {\rm v}^2-n^2 {\rm v}^1)+g^{11}n^3(n^2 {\rm v}^3-n^3 {\rm v}^2)],\\
&&Z\equiv(\varepsilon_{123})^2[g^{22}n^1(-n^1 {\rm v}^3+n^3
{\rm v}^1)-g^{11}n^2(n^2 {\rm v}^3-n^3 {\rm v}^2)]\ ,
\end{eqnarray}
where the components $n^i$, ${\rm v}^i$ are computed in the curved
space time and
$\varepsilon_{123}=\frac{r^2\sin{\theta}}{\sqrt{(1-r_S/r)}}$.
Using (\ref{a32}), (\ref{a33}) and their derivatives,
we find:\newline\
\begin{eqnarray}\label{se22a}
X_0&=&2lN\sin{(2Nt)} \left\{\cos^2{\lambda} \, e(1-e^2) \times \left[1-2\mu(1+e^2)-(1-4\mu)e\cos{(2 N t)} \right]^{1/2} \right .
\nonumber\\
&& \times \left[1-e\cos{(2Nt)} \right]^{-1/2} \left[(1+e^2)-2e\cos{(2N t)} \right]^{-2} \nonumber \\
&&-\sin^2{\lambda}(1-e^2) \left[\cos{(2Nt)}-e \right]  \nonumber\\
&& \times[1-2\mu(1+e^2)-(1-4\mu)e\cos{(2 N t)}]^{1/2}] \nonumber \\
&& \left . \times[1-e\cos{(2Nt)}]^{-3/2}
[(1+e^2)-2e\cos{(2N t)}]^{-2} \right \},
\end{eqnarray}
\begin{eqnarray}
\frac{1}{r}Y_0&=&-2lN\sin{\lambda}\cos{\lambda} \frac{\sqrt{(1-e^2)}[(1+e^2)\cos{(2Nt)}-2e]}{[(1+e^2)-2e\cos{(2Nt)}]^2}, \label{se22ax}
\end{eqnarray}
\begin{eqnarray}
\frac{1}{r}Z_0&=&-2lN \left\{\sin^2{\lambda}
\frac{(1-e^2)^{3/2}\sin^2{(2Nt)}}{[1-e\cos{(2Nt)}][(1+e^2)-2e\cos{(2N t)}]^2} \right .\nonumber \\
&& \left . +\cos^2{\lambda}\frac{\sqrt{1-e^2}}{[(1+e^2)-2e\cos{(2Nt)}]} \right\},
\label{se22ay}
\end{eqnarray}
while the term
$\frac{r\sin{\lambda}\sin{\phi}}{\sqrt{(1-r_S/r)}}$ in (\ref{se21b}) becomes
\begin{eqnarray}\label{se22az}
\frac{r\sin{\lambda}\sin{\phi}}{\sqrt{(1-r_S/r)}}&=&l\sqrt{1-e^2}\sin{\lambda}\sin{(2Nt)}
\times \left[1-e\cos{(2Nt)} \right]^{-3/2} \nonumber \\
&&\times \left [1-2\mu(1+e^2)-(1-4\mu)e\cos{(2Nt)} \right ]^{-1/2}.
\end{eqnarray}
Keeping terms up to $e^2$, an approximate form of the
(\ref{se22a}-\ref{se22az}) is
\begin{eqnarray}\label{sa1}
X_0&=&2lN\sin{(2Nt)}\cos^2{\lambda}[(1-2\mu)^{1/4} e+O(e^2)],\\
\frac{1}{r}Y_0&=&-2lN\sin{\lambda}\cos{\lambda}
\times\{\cos{(2Nt)}+2[\cos^2{(2Nt)}-1]e+O(e^2)\},\\
\frac{1}{r}Z_0&=&-2lN\{\sin^2{\lambda}\sin^2{(2Nt)}
\times[1+5\cos{(2Nt)}e+O(e^2)\},
\end{eqnarray}
and
\begin{eqnarray}\label{sa4}
{\bf n}\cdot {\bf r}=l\sin{\lambda}\sin{(2Nt)}\left[\frac{1}{\sqrt{1-2\mu}}
+\frac{(2-5\mu)}{(1-2\mu)^{3/2}}\cos{(2Nt)}e+O(e^2)\right] \, .
\end{eqnarray}
\section*{Appendix B}

In (\ref{f1}) we computed the ratio $L/E$ with the aid of (\ref{e6}), (\ref{latusrectum}). In this case, we find
\begin{eqnarray}\label{b1}
\frac{E^2}{L^2}&=&\frac{2}{r_S l}\left[1-(3+e^2)(\mu-\frac{\Omega}{l})\right]
-\frac{1}{l^2}\left[1-4(\mu-\frac{\Omega}{l})(1-e^2)\right].
\end{eqnarray}
Furthermore, (\ref{b1}) gives
\begin{eqnarray}\label{b2}
\frac{L}{E}&=&\pm \sqrt{M l}\left\{(1-2\mu)^2-4\mu^2 e^2 +\frac{\Omega}{l}[(3+e^2)-4\mu(1-e^2)]\right\}^{-1/2}.
\end{eqnarray}
In (\ref{b2}), setting $\Omega=0$ and $e=0$, we recover
Breuer's result ~\cite{a7}. In the case
where $\Omega=0$ and $e\neq 0$, we recover the result
obtained in  \cite{a5y}.

From (\ref{f1}) and (\ref{b1}) we find
\begin{eqnarray}\label{b4}
2N&=&\pm \frac{M^{1/2}}{l^{3/2}}\left\{(1-2\mu)^2-4\mu^2 e^2
+\frac{\Omega}{l}[(3+e^2)-4\mu(1-e^2)]\right\}^{-1/2} \nonumber \\
&\times&\left[1+\frac{\Omega}{l}(1-2e)\right]^{-1} \times\left[1-(3-e)(2\mu-2\frac{\Omega}{l})\right]^{1/2}.
\end{eqnarray}
In the limit $\Omega=0$, we recover the coefficient
of the Eq. (136) in~\cite{a5y}.
Observe, that on the right side of (\ref{b4}) the coefficient $\pm{{M}^{1/2}}/{l^{3/2}}$
corresponds to the Newtonian frequency of a Keplerian orbit at $r=l$ in $c=G=1$ units.


\end{document}